\documentclass[conference]{IEEEtran}
\ifCLASSINFOpdf
\else
\fi
\usepackage{amsmath,amssymb}
\usepackage{graphicx}

\usepackage[utf8x]{inputenc}

\usepackage{textcomp,marvosym}

\usepackage{cite}

\usepackage{nameref,hyperref}

\usepackage[right]{lineno}

\usepackage[table]{xcolor}

\usepackage{array}
\newtheorem{theorem}{Theorem}[section]

\usepackage{siunitx}
\usepackage{changepage}

\begin{document}
%
\title{The spread of COVID-19 at Hot-Temperature Places With
Different Curfew Situations Using Copula Models}

\author{\IEEEauthorblockN{Fadhah Alanazi}
\IEEEauthorblockA{Deanship of Educational Services,\\ Department of General Sciences,\\ Prince Sultan University,\\ Riyadh, Saudi Arabia\\
Email:fanazi@psu.edu.sa}
}


%


\maketitle
\section*{Abstract}
The infectious coronavirus disease 2019 (COVID-19) has become a serious global pandemic. Different studies have shown that increasing temperature can play a 
crucial role in the spread of the virus.  Most of these studies were limited to winter or moderate temperature levels and were conducted using conventional models. However, traditional models 
are too simplistic to investigate complex, non-linear relationships and suffer from some restrictions. Therefore, we employed copula models to examine the impact 
of high temperatures on virus transmission. The findings from the copula
models showed that there was a weak to moderate effect of temperature on the
number of infections and the effect almost vanished under a lockdown policy.
Therefore, this study provides new insight
into the relationship between COVID-19 and temperature, both with and without
social isolation practices. Such results can lead to improvements in our
understanding of this new virus. In particular, the results derived from the
copula models examined here, unlike existing traditional models, provide
evidence that there is no substantial influence of high temperatures on the
active COVID-19 outbreak situation. In 
addition, the results indicate that the transmission of COVID-19 is strongly
influenced by social isolation practices. 
To the best of the authors’ knowledge, this is the first copula model
investigation applied to the COVID-19 pandemic.

%

\section*{Introduction}
In December 2019, a novel infectious disease termed coronavirus disease 2019 (COVID-19) 
was discovered in Wuhan city, Hubei province, China. 
Subsequently, and through human-to human transmission, this virus has 
caused a global pandemic. 
COVID-19 is characterized by clinical features similar to those caused by 
severe acute respiratory syndrome coronavirus (SARS-CoV) and Middle 
Eastern respiratory syndrome coronavirus (MERS-CoV) infections, 
such as a fever and dry cough \cite{huang20}. \\~\\
Previous studies have shown that meteorological variables can
 affect the transmission and survival of coronaviruses 
 \cite{SARS, alghamdi2014}. 
Earlier research \cite{alghamdi2014} found that MERS-CoV is
 most active at high temperatures and low humidity. \\~\\
Notably, recent studies have shown that warm weather and high humidity 
may be important factors for reducing the spread of COVID-19 
(e.g., see \cite{Wang2020}). 
Conversely, some existing studies have found that increasing
 temperatures will not affect the transmission of COVID-19
 (e.g., see \cite{XIE20}).
However, most of these studies were limited to winter or
 low-temperature weather with a small number of
  observations. Hence, there is still no definitive
evidence as to whether there is a negative association between environmental variables and the spread of COVID-19 in extremely hot or cold locations \cite{Wang2020}.
Besides, most previous studies were performed using traditional models, which are too simplistic and may be unable to deal with complex, non-linear dependency 
patterns. Thus, further research to understand the activity of COVID-19 under high-temperature conditions is warranted. In addition, such an association should be investigated 
not only in regard to weather variables, but also by taking into account the lockdown situation at these locations. 
Presently, copula models have become a favored statistical tool to describe the association between variables. These models have been applied 
in different areas, including the study of infectious diseases (e.g., see \cite{ray2017infectious}) and environmental science (e.g., see \cite{grandstrand:2004}). 
One important benefit of using a copula model is that one can model the marginal distribution independently from the dependency structures, which are completely captured via the copula 
function. Another benefit of using a copula model is that the margins do not need to follow the same parametric family. Furthermore, many copula families exist, 
each with its own capability to describe the unique dependency structure. Hence, various types of associations can be discovered via copula models.
\\~\\
Hence, this study aimed to perform flexible statistical modeling with a copula model to improve our knowledge about the spread of the virus in hot locations with different curfew levels. Specifically, we investigated the impact of high temperatures on the number of confirmed cases in the cities of Riyadh, Jeddah, and 
Mecca in Saudi Arabia, and these cities were selected for several reasons. First, Saudi Arabia has been strongly affected by MERS-CoV  
\cite{alshukairi2018, MERCOV}, which produces a similar severe respiratory illness as COVID-19. Second, the highest numbers of confirmed cases in Saudi Arabia
have been recorded in Riyadh, Jeddah, and Mecca, which are three of the hottest areas in Saudi Arabia. Third, because of the transmission of COVID-19,
Mecca and Jeddah have been placed under a series of lockdowns for a long time. Riyadh, however, was only placed under a curfew for a short period.
Hence, these cities represent strong to moderate lockdown situations, which could be a factor critical to understanding the effects of high temperatures 
on the spread of COVID-19. By using the data from these cities and capitalizing on the flexibility of copula models, we aimed to provide clear evidence on the association between high temperatures and 
confirmed cases of COVID-19.

\section*{Materials and methods}

\subsection*{Data collection}
For this study, the cities of Riyadh, Mecca, and Jeddah were selected for the analysis. Riyadh is  the capital city of Saudi Arabia and the city
most affected by COVID-19 in this country. Riyadh had $37,244$ confirmed cases for the observed period from $13$ March $2020$ to $15$ June $2020$. 
The population of Riyadh estimated for the middle of 2018 based on demographic survey data collected in 2016 was $8,446,866$ \cite{Riyadh2018}. Jedda and Mecca are the second
most affected cities in Saudi Arabia, with $21,152$ and $20,248$ confirmed cases, respectively. \\
Daily counts of confirmed cases for the study period were collected from official reports \cite{KAB} based on information 
\cite{MOH} for Saudi Arabia. The daily average temperature data for the same period time were obtained 
from the Weather Underground Company \cite{weather}. 
\subsection*{Data analysis}
Both the COVID-19 confirmed cases and the average temperature data demonstrated a non-normal distribution for all cities. 
Fig 1 \ref{CaseR} shows the confirmed cases during the study period, where the new confirmed cases of COVID-19 in Riyadh exceeded $1500$ from $08$
to $15$ June $2020$. However, the highest records for Mecca and Jeddah were generally similar and lower than those of Riyadh. 

\begin{figure}[!h]
 \includegraphics[scale=0.28]{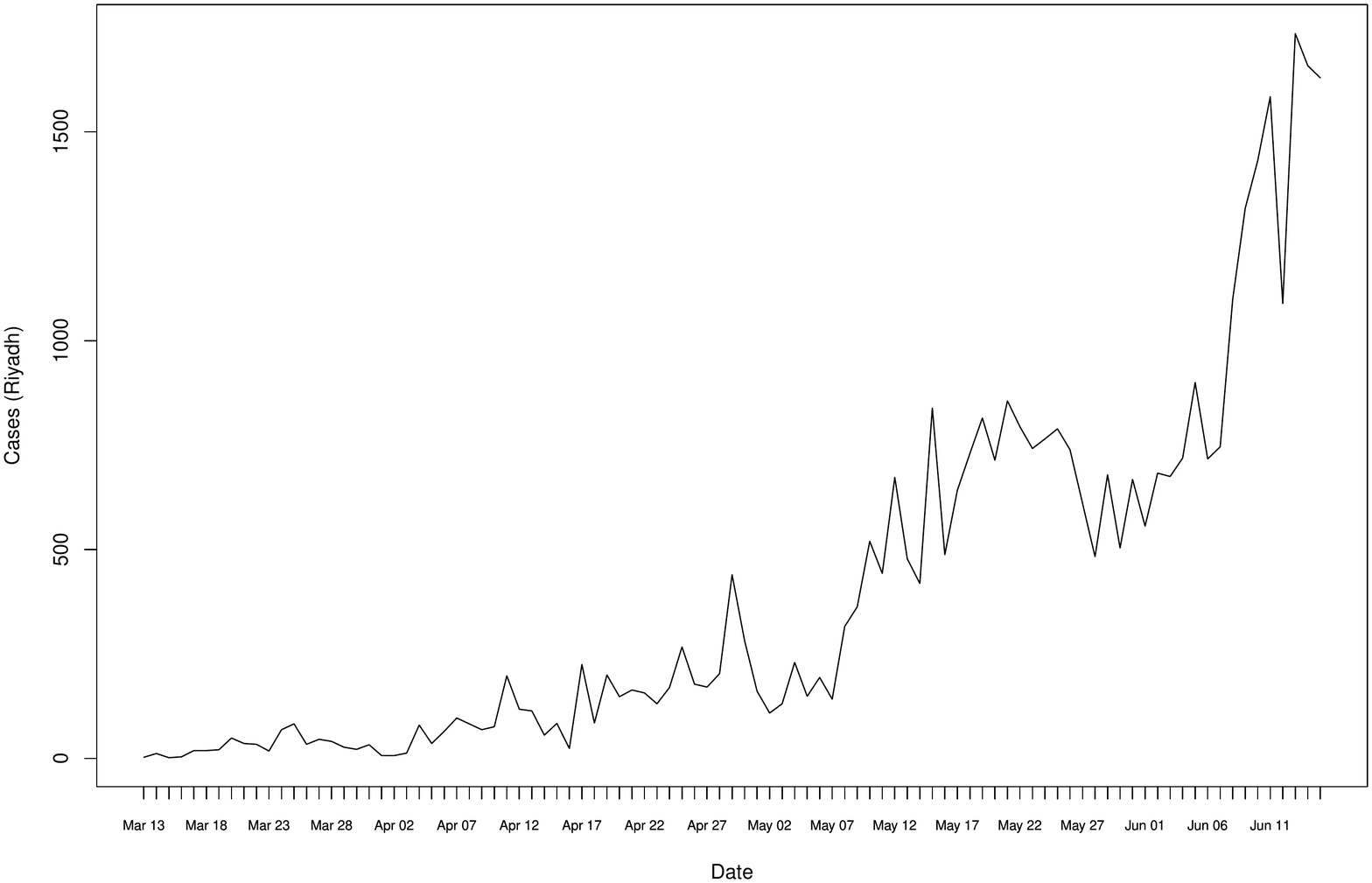} \\
 \includegraphics[scale=0.28]{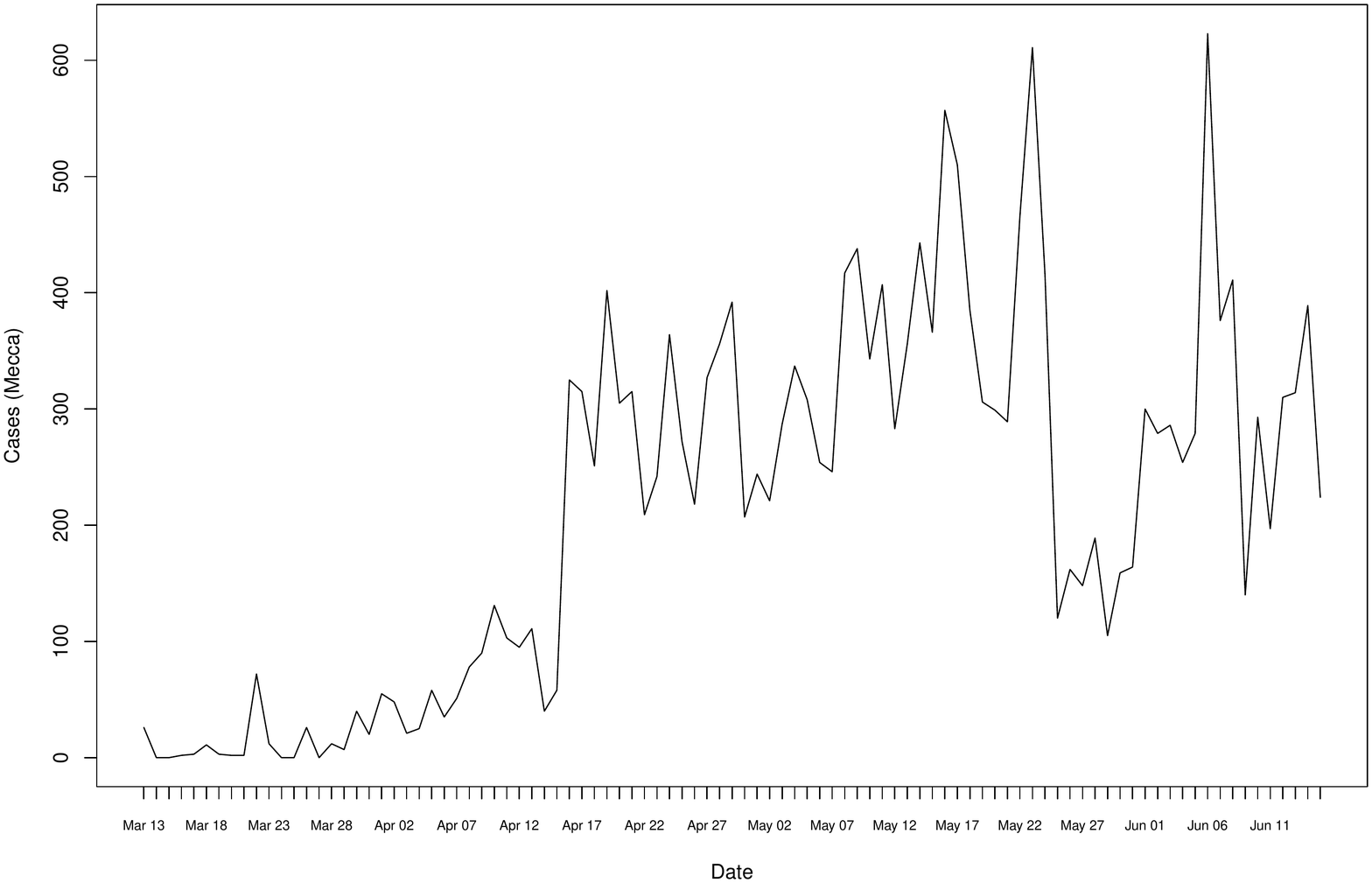} \\
\includegraphics[scale=0.28]{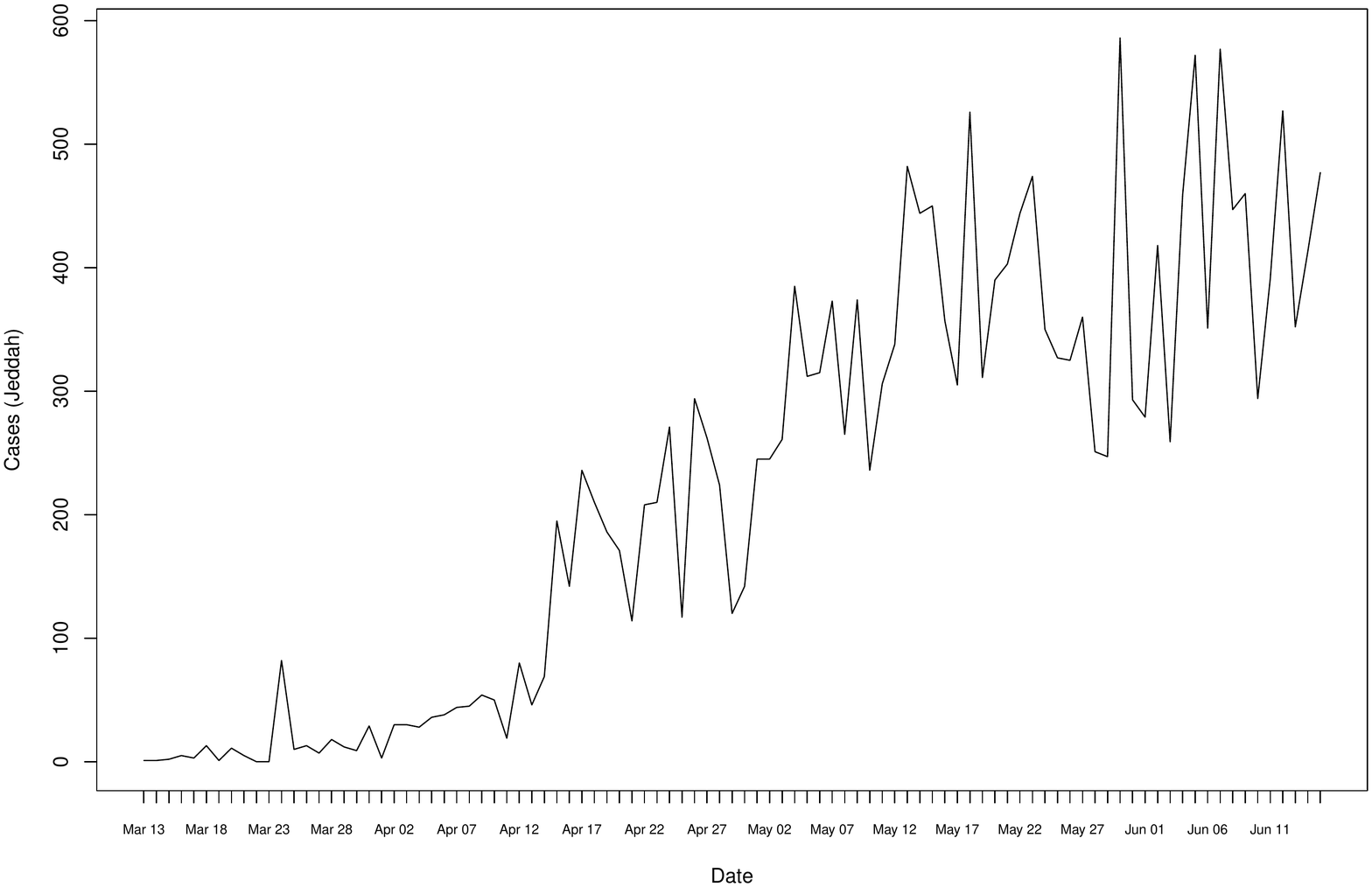}
\caption{{\bf Plots of the confirmed cases during the study period (from $13$ March $2020$ to $15$ June $2020$.}
(top): Riyadh, (middle): Mecca, (bottom) Jeddah.}
\label{fig1}
\end{figure} \label{CaseR}
\subsection*{Copula}
Copula is a Latin word that means joins or links. A copula function refers to a multivariate function that joins the multivariate distribution functions to their
univariate standard uniform margins \cite{Nelsen2006}.
Formally, a copula can be defined as follows:
copulas \cite{Schweizer2011} are multivariate cumulative distribution functions with uniform marginal distributions on (0,1) such that:
\begin{equation}
              C:[0,1]^n \rightarrow [0,1], \ \ \ \  \  n\geq 2.
\end{equation}
Sklar’s theorem \cite{Sklar1959} is the key rule of the copula function, and it can be introduced as follows:
\\
\begin{theorem}[Sklar's theorem]
If $ F $ is an $n$-variate distribution function with univariate margins $ F_1, F_2,
.....,F_n $, then there exists
an $n$-variate copula function, $ C $, such that $\forall$ $\mathbf{x}=(x_1, ..,x_n)^{'} \in \mathcal{R}^{n}$:
\begin{equation}\label{sklar'sequation}
F(x_1,x_2,....,x_n)= C(F_1(x_1),F_2(x_2),....,F_n(x_n)).
\end{equation}
If the margins are continuous, then the copula
\begin{equation}
C(u_1, u_2, ....,u_n)= F(F^{-1}_{1}(u_1), F^{-1}_{2}(u_2),...,F^{-1}_{n}(u_n))
\end{equation}
is unique, where $F^{-1}$ is the inverse function of the margins and  $u \in [0,1]^{n}$. 
Conversely, if $F_1, ..., F_n$  are the
marginal distribution functions and $C$ is a copula 
function, then the function $F$ (defined by equation 
(\ref{sklar'sequation})) is a joint distribution function with margins $F_1, ..., F_n$.
\end{theorem}
In accordance with Sklar's theorem (\ref{sklar'sequation}), a copula models the marginal distributions separately from the dependency pattern, with no restriction
on the type of margins. 
\\~\\
In this study, we consider an arbitrary number of copula types including the Joe, Gumbel, and Clayton copulas, as well as their rotation types. 
In addition, we consider the Frank, Gaussian, t-students, and other two-parametric copulas, such as the Joe-Frank (BB8) copula. The following text provides
details on some commonly used copula families. 
\begin{itemize}
	\item \textbf{Frank copula} is a one-parametric symmetric Archimedean copula with generator function 
    $\varphi(t) = - \ln
    [\frac{e^{-\theta t} -1 }{e^{-\theta} -1}]$, with $\theta \in (- \infty, \infty)$ $\backslash$ $\{0\}$. 
 The Frank copula can control both the negative and positive dependency pattern, where the strongest dependency occurs at the center of the distribution. 
 However, in the Frank copula, the extremes are independent. \\~\\
    The distribution function of the Frank copula can be given by:
   \begin{equation}
        C_{\theta}(u_1, u_2) = \frac{-1}{\theta} \ \ln  \ [1 + \frac{(e^{-\theta u_1}-1)(e^{-\theta
       u_2}-1)}{e^{-\theta} -1}],
   \end{equation}
   and its density function is:
   \begin{equation}
       c(u_1, u_2)= \theta (e^{-\theta}-1) \frac{e^{-\theta}(u_1+u_2)}{e^{-\theta}-1+(e^{-\theta u_1}
       -1)(e^{-\theta u_2} -1)}.
  \end{equation}
  \item \textbf{Clayton copula} is a one-parametric ($ \theta > 0$) non-symmetric Archimedean copula. It is a lower positive tail dependence copula 
    with generator $\varphi(t)= \frac{1}{\theta}(t^{-\theta}-1)$.
   Its distribution is given by:
       \begin{equation}
        C(u_1, u_2) = [u_1^{-\theta}+u_2^{-\theta}-1]^{\frac{-1}{\theta}},
        \end{equation}
        and its density function is:
        \begin{equation}
        c(u_1, u_2)= (1+\theta)(u_1 u_2)^{-1-\theta} (u_1^{-\theta}+u_2^{-\theta} -1)^{\frac{-1}{\theta} -2}.
    \end{equation}
 \item \textbf{Joe copula}, in contrast to the Clayton copula, this is a one-parametric upper tail Archimedean copula with generator
    $\varphi(t)=
    \ln[1-(1-t)^{\theta}]$. 
     Its distribution function is:
    \begin{equation}
        C(u_1, u_2)= 1- [(1-u_1)^{\theta}+(1-u_2)^{\theta}-(1-u_1)^{\theta}(1-u_2)^{\theta}]^{\frac{1}{\theta}},
    \end{equation}
    and its density function is:
    \begin{equation}
    \begin{split}
    c(u_1,u_2)=[(1-u_1)^{\theta}+(1-u_2)^{\theta}-(1-u_1)^{\theta}
    (1-u_2)^{\theta}]^{\frac{1}{\theta} -2}\\
    \times (1-u_1)^{\theta-1}(1-u_2)^{\theta-1}
  [\theta -1+(1-u_1)^{\theta}\\
  +(1-u_2)^{\theta}-(1-u_1)^{\theta}(1-u_2)^{\theta}].
    \end{split}
    \end{equation} 
    \item \textbf{Rotated copula} refers to a rotation version of asymmetric
copulas. This rotation includes $90$, $180$, and $270$ rotation degrees, with
arguments $(1-u_1, u_2), (u_1, 1-u_2)$, and $(1-u_1, 1-u_2)$,
respectively. The $180$ rotation degree produces a corresponding survival copula family. However, rotations by $90$ and $270$ degrees provide
corresponding copulas to deal with negative
dependencies. For more details on rotated copulas, see for example, 
\cite{cech2006copula}, \cite{wiboonpongse2015modeling}, \cite{luo2011stepwise},
 and \cite{Dissmann2013}.
\end{itemize}
\subsubsection*{Pseudo maximum-likelihood method}
In this study, we applied the so-called \textit{pseudo maximum-likelihood method} (\textit{PML}) to estimate the parameters for the selected copula function. 
\textit{PML} is introduced by \cite{Gen1995} as a two-step estimation
method. With this method, the margins are estimated non-parametrically via their empirical cumulative distribution function at
first, and then, the copula parameter ($\theta_c$) is estimated at the second step. By using \textit{PML}, the copula parameter is
estimated by maximizing the copula density, i.e.,
\begin{equation}
    L_{\textbf{MPL}} (\theta_c) = \sum_{i=1}^{n} \log [c(u_{1i},u_{2i};\theta_c)],
\end{equation}
where $u_1 = \hat{F}_1(x_1;\alpha_1)$ and $u_2= \hat{F}_2(x_2;\alpha_2)$ are the empirical probability integral transform of variable $X_1$ and $X_2$, 
respectively. A simulation study of \cite{KIM20072836}
showed that the performance of \textit{PML} is better than that of the full maximum likelihood estimation method and  
\textit{Inference Function of Margins} of \cite{Joe1997} if the margins are unknown, which is the case in almost all real life 
applications.
\subsection{Goodness-of-fit test}
As there is a wide range of copula functions, it is necessary to test the copula shape with the best fit. Therefore,
we will use the Akaike Information Criterion (\texttt{AIC}) of \cite{Akaike1973} and
the Bayesian Information Criterion (\texttt{BIC}) of \cite{schwarz1978estimating} to select the right copula. 
\texttt{AIC} and \texttt{BIC} can be given by:
 \begin{equation}
\texttt{AIC} = -2 \ \texttt{ln} \ \texttt{L}(\hat{\theta}) + 2 \* \texttt{P},
\end{equation}
 \begin{equation}
 \texttt{BIC} = -2 \ \texttt{ln}  \ \texttt{L}(\hat{\theta}) +  \texttt{P} \ (\texttt{ln (N))},
\end{equation}
where $\hat{\theta}$ is the estimated value of the parameters, and \texttt{P} is the number of the model parameters.
\\~\\
The summary of  the full inference steps of copula models used in this study is as follows:
\begin{itemize}
	\item Transform the continuous variable of the observed data to copula data. 
	\item Calculate the cumulative density function for the discrete variable of the observed data. 
	\item Consider arbitrary types of bivariate copula functions for the assumed model.
	\item Select the best fit bivariate copula type among all fitted copula functions using \texttt{AIC} and \texttt{BIC}.
\end{itemize}

\section*{Results and discussion}
\subsection*{Descriptive results}
Table \ref{DesT} shows the summary statistics for the daily data on temperature and COVID-19 confirmed cases in the cities of Riyadh, Mecca, and Jeddah. 
With average values of \ang{29.788}C (Riyadh), \ang{24}C (Mecca), and \ang{29.014}C (Jeddah), the temperature in the three cities was very high. Importantly, 
this study takes into account the daily average temperature, and not the maximum temperature.

\begin{table}[!ht]
\centering
\caption{
{\bf Summary statistics for the daily average temperature and COVID-19 confirmed cases in the cities of Riyadh, Mecca, and Jeddah. Zeros values
  indicate no new confirmed cases for the corresponding date during the observation period.}}\label{DesT} 
\begin{tabular}{@{\extracolsep{0.8pt}} |l|l  |l | l |  l| l|} \hline 
Variable & N & Mean & Standard deviation & Min & Max \\ 
\hline 
Temperature (Riyadh) (\textdegree C) & $95$ & $29.788$ & $4.947$ & $18$ & $37.4$ \\
 \hline
Temperature (Mecca) (\textdegree C) & 95 & 24.983 & 3.472 & 18  & 32 \\ 
\hline
Temperature (Jeddah) (\textdegree C) & 95 & 29.014 & 3.258 & 22.9 & 36 \\ 
\hline
Confirm cases (Riyadh) & 95 & 392.042 & 427.393 & 2 &  1,735 \\
 \hline
Confirm cases (Mecca) & 95 & 213.137 & 160.938 & 0  & 623 \\ 
\hline
Confirm cases (Jeddah) & 95 & 222.653 & 174.632 & 0 & 586 \\ 
\hline 
\end{tabular} 
\end{table}

\subsection*{Discussion of the Copula model}
\begin{figure}[!ht]
\includegraphics[scale=0.3]{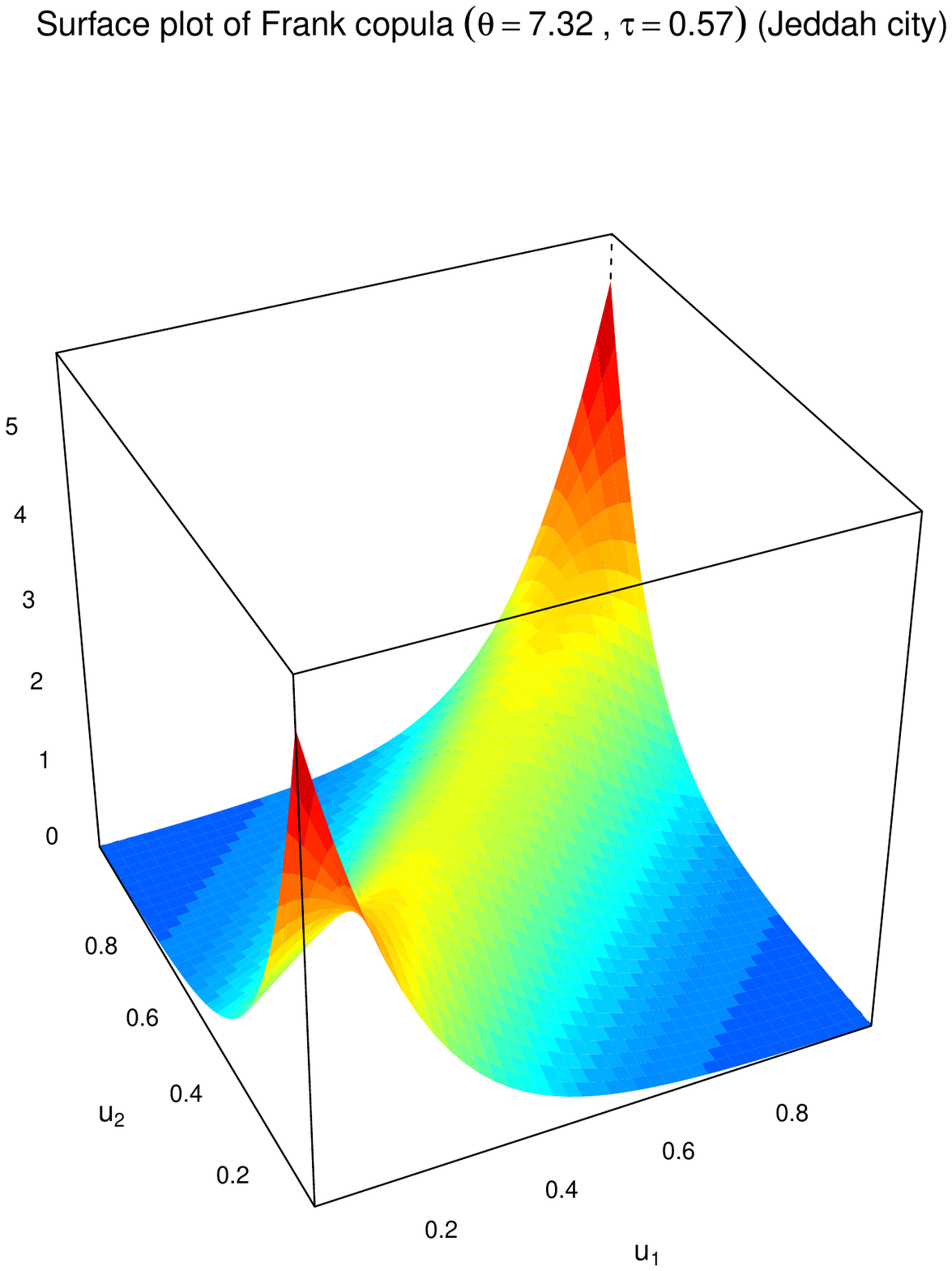} \includegraphics[scale=0.3]{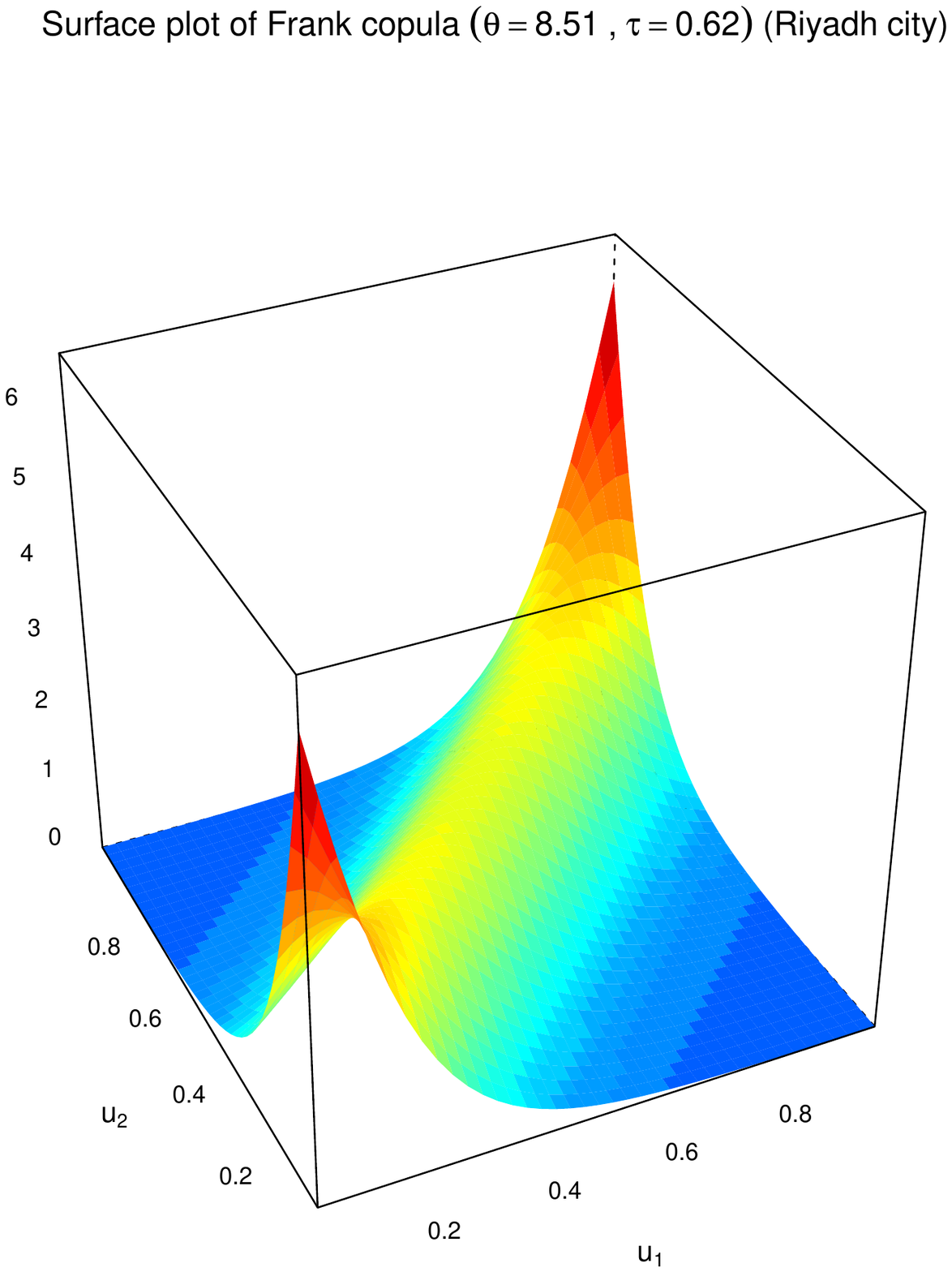}
\includegraphics[scale=0.3]{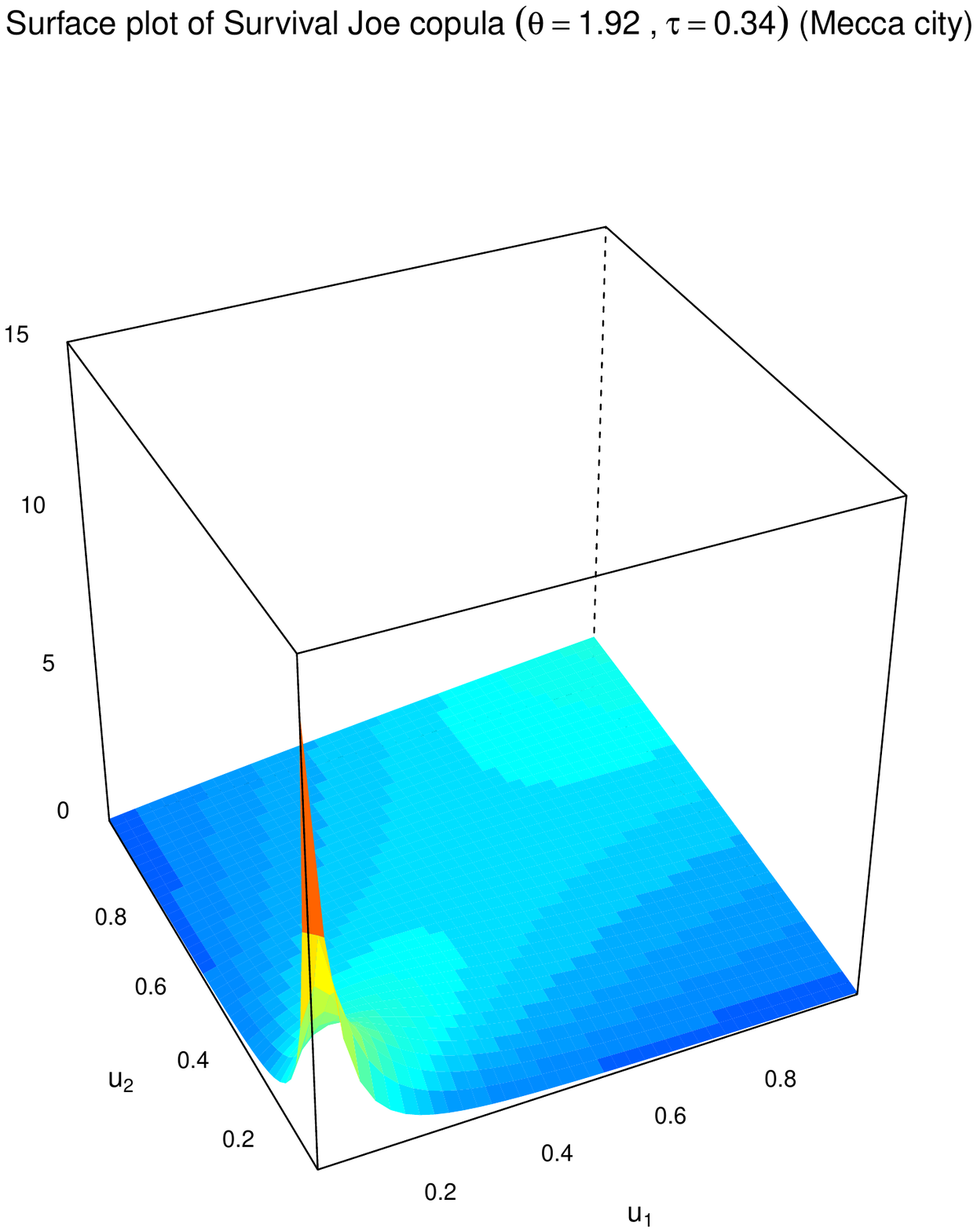}
\caption{{\bf Surface plots of the Frank copulas for Riyadh (top) and Jeddah (middle), and the Clayton copula for Mecca (Bottom) as the best-fit copula families for each city.} }
\label{Surface}
\end{figure}

\begin{figure}[!ht]
\includegraphics[scale=0.22]{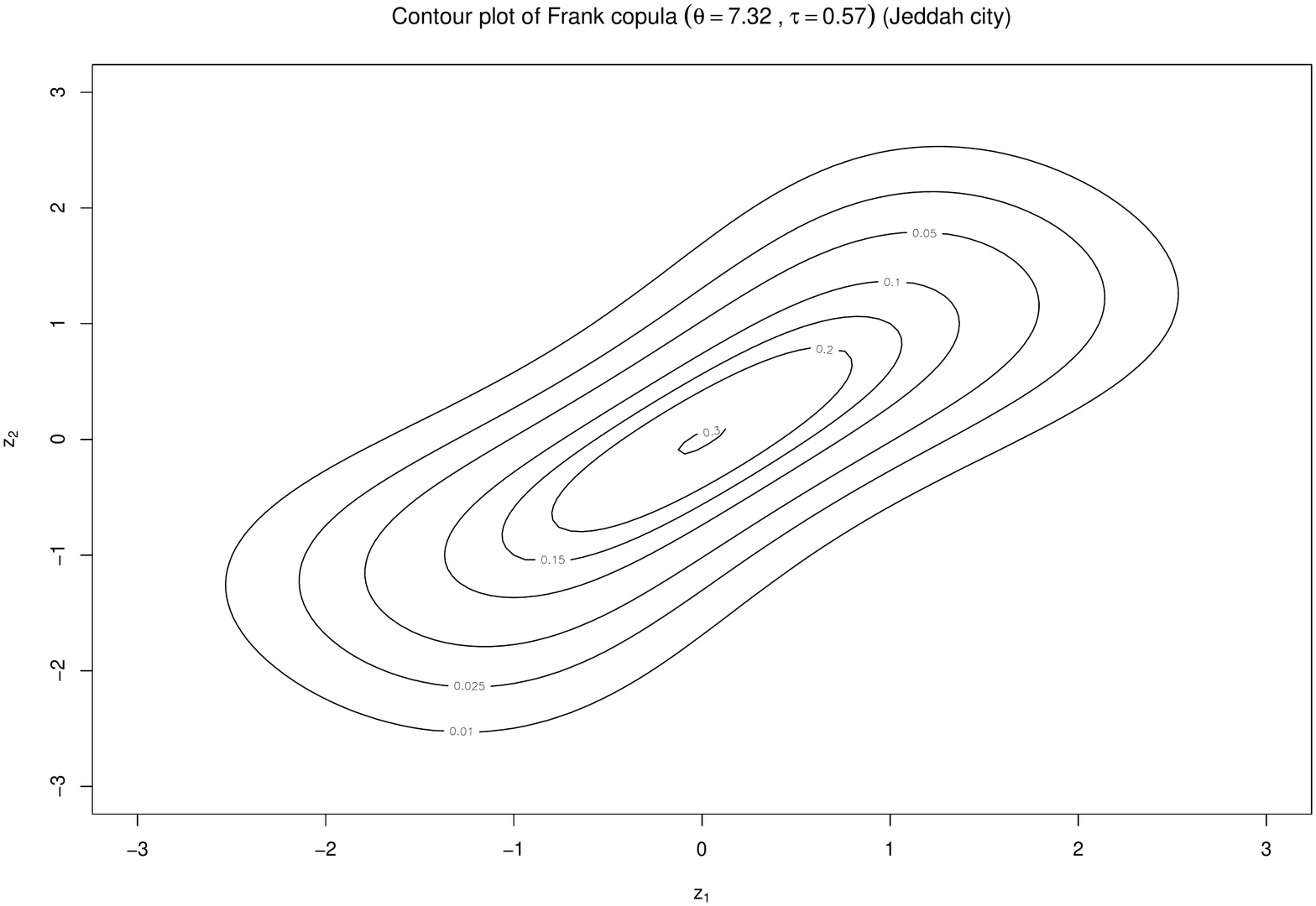}
\includegraphics[scale=0.22]{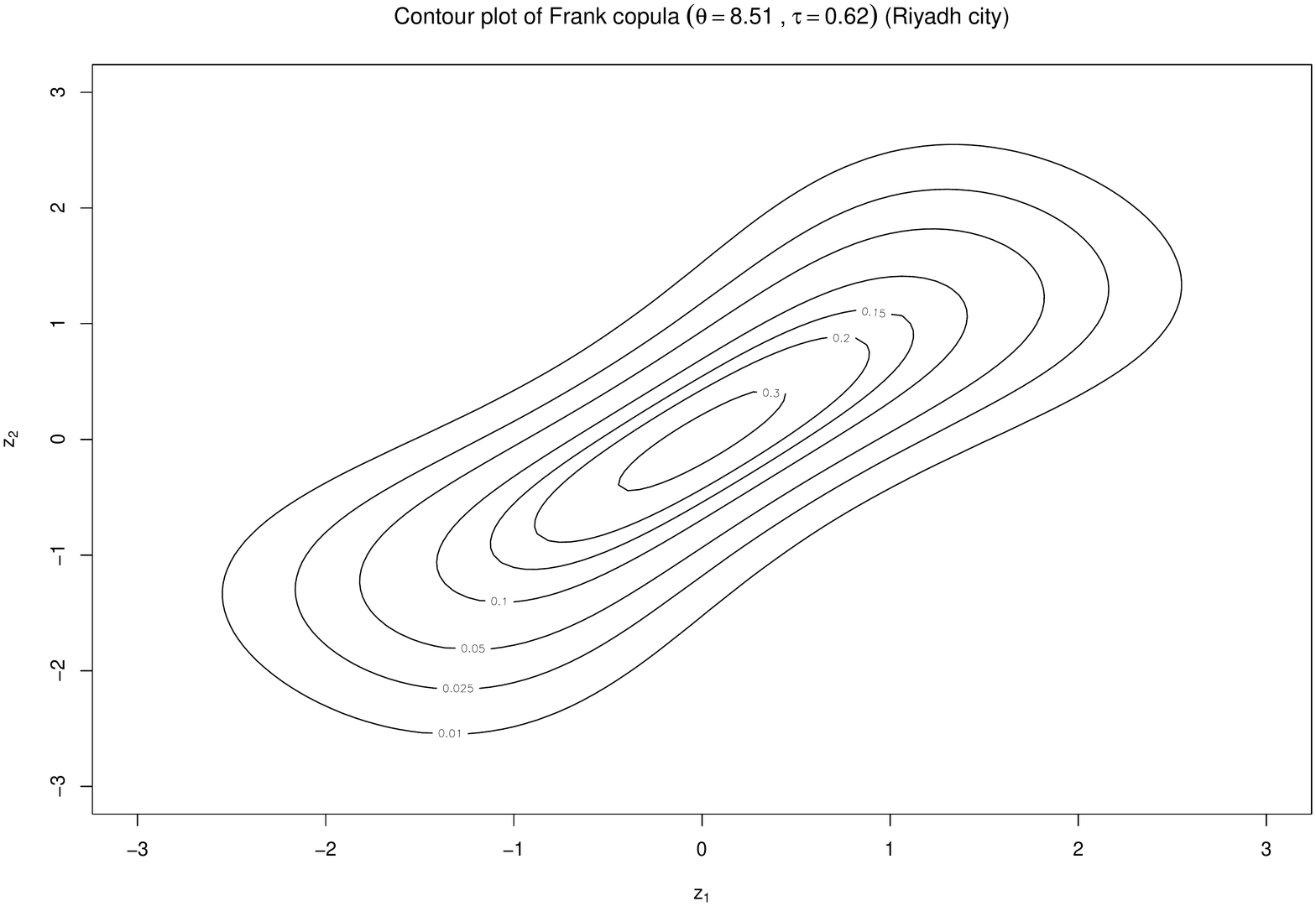} 
\includegraphics[scale=0.22]{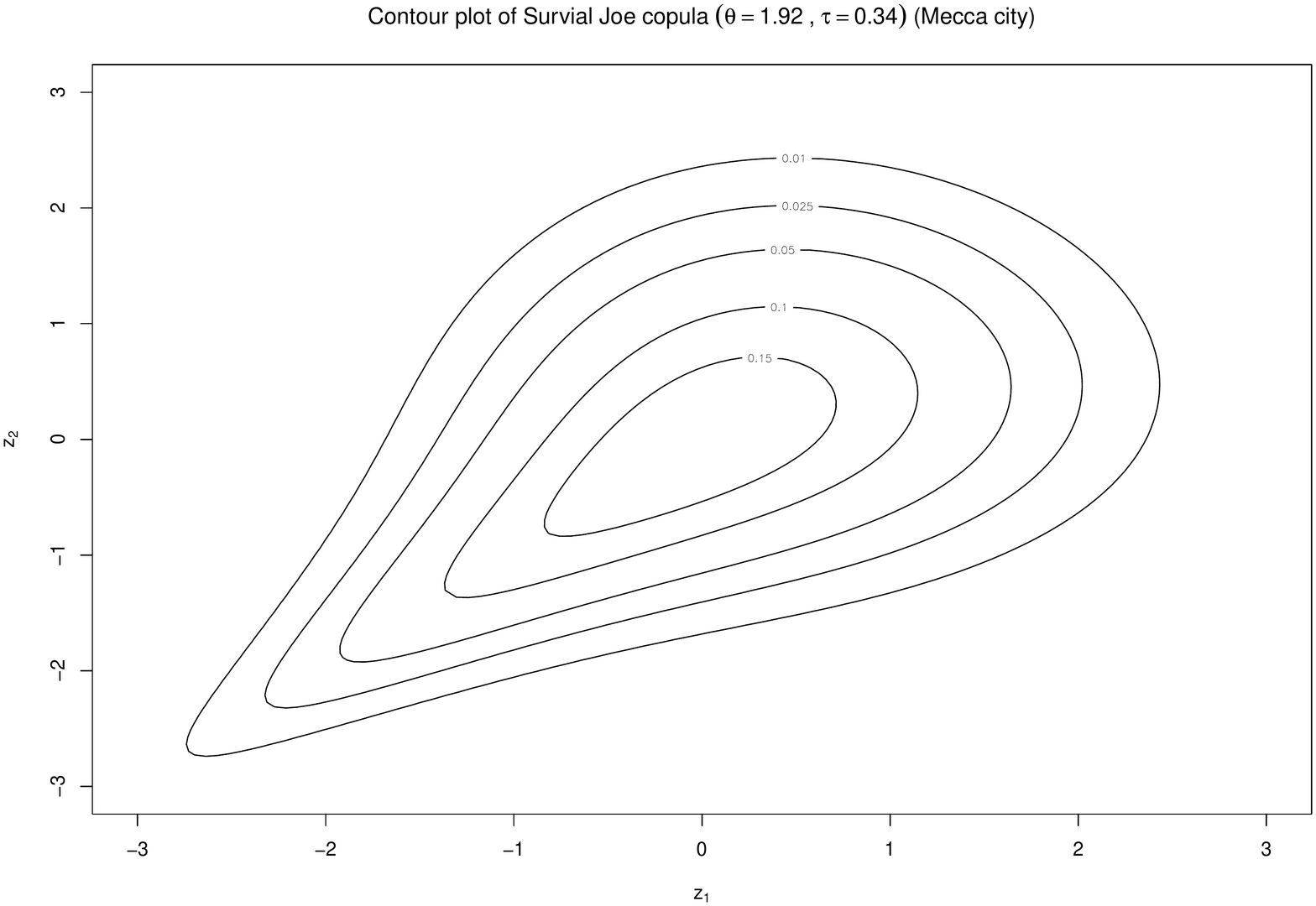}
\caption{{\bf Contour plots of the Frank copulas for Riyadh (top) and Jeddah (middle), and the Clayton copula for Mecca (Bottom) as the best-fit copula families for each city.} }
\label{Contour}
\end{figure}

This study used a copula model to investigate the relationship between high temperatures and confirmed cases of COVID-19 for three cities in Saudi Arabia.
The existing number of copulas is large, and to select the most appropriate fitted model for each city, \texttt{bicop()} function of \texttt{R} (\cite{Rprogram})
package \cite{rvinecopulib} was used. As \texttt{bicop()} allows one to consider different selection criteria at each run, it was applied to each data set twice,
one with \texttt{BIC} and the other one with  \texttt{AIC}. The results of these models are provided in 
Tables (\ref{BIC_c}, \ref{AIC_c}). \textit{Boldfont} indicates the selected copula type. Figs (\ref{Surface}, \ref{Contour}) present the surface and 
contour plots for the selected copula families.  \\~\\

\begin{table}[!ht] \centering 
\caption{ \bf Selected copula families (for each city) based on BIC by \texttt{bicop()}} 
  \label{BIC_c} 
\begin{tabular}{@{\extracolsep{0.2pt}} |l|l| l| l| } \hline
Type & Parameter ($\tau$) & AIC & BIC \\ 
\hline 
& Riyadh & & \\
\hline 
Frank & $\theta = 8.52$ ($\tau = 0.62$) &   $\mathbf{-102.5}$ &   $\mathbf{-99.97}$ \\
\hline 
& Mecca & & \\
\hline 
Survival Joe (rotation degree $180$) & $\theta = 1.92$ ($\tau =0 .43$) &   $\mathbf{-35.07}$ &   $\mathbf{-32.51}$  \\    
\hline 
& Jeddah & & \\
\hline 
Frank & $\theta = 7.32$ ($\tau =0 .57$) &   $\mathbf{-85.52}$ &  $\mathbf{-82.97}$ \\  
\hline
\end{tabular} 
\end{table} 

\begin{table}[!h]
\begin{adjustwidth}{-0.25in}{0in} 
\centering
\caption{\bf Selected copula families (for each city) based on AIC by \texttt{bicop()}.} 
  \label{AIC_c} 
\begin{tabular}{@{\extracolsep{3pt}} |l|l| l| l| } \hline
Type & Parameter ($\tau$) & AIC & BIC \\ 
\hline 
& Riyadh & & \\
\hline 
BB8 (Joe-Frank)  & $\theta = 8$ ($\alpha = 0.7$) & $-103.74$ &  $-98.63$ \\
\hline 
& Mecca & & \\
\hline 
tll \footnote{(nonparametric (transformed kernel))} & $7.47$ df \footnote{(df = degree of freedom)}&  $-39.32$ & $-20.24$ \\    
\hline 
& Jeddah & & \\
\hline 
Frank & $\theta = 7.32$ ($\tau =0 .57$) &  $\mathbf{-85.52}$ &  $\mathbf{-82.97}$ \\  
\hline
\end{tabular} 
\end{adjustwidth}
\end{table} 
\newpage
In accordance with Table (\ref{BIC_c}), Frank copulas with a moderate positive dependency ($\theta = 8.51$, $\tau = 0.62$) and (($\theta = 7.47$, $\tau = 0.57$) were 
selected for Riyadh and Jeddah, respectively. The results indicate that there is a positive relationship between temperature and the spread of COVID-19
in moderate and high temperatures. However, these two variables became independent at extreme values. 
Therefore, the results provide clear evidence that SARS-CoV-2 can still remain an active virus in hot places. 
In the case of Mecca, the survival Joe copula with a low dependency level ($\theta = 1.92$, $\tau= 0.34$) 
was selected as the most appropriate copula function. With these data, there was only a very weak dependency pattern between high temperatures 
and confirmed cases at the low values. This relationship reflects the period before the series of lockdowns in the city of Mecca. During the curfew, there 
was no relationship detected between the spread of COVID-19 and the temperature.
\\~\\
The question that now remains is, do high temperatures affect the transmission of COVID-19? To answer this question, we need to mention some of the main similarities 
and differences among these cities regarding the (1) temperature, (2) confirmed cases of COVID-19, (3) copula model results, and (4) lockdown 
situations.
First, the temperatures in Riyadh and Jeddah were almost the same and slightly higher than those in Mecca. However, the number of confirmed cases in 
Riyadh was higher than that in Jeddah and Mecca. In addition, Mecca and Jeddah had almost the same number of confirmed cases. Hence, the same temperature levels 
were associated with different numbers of new confirmed cases within the three cities. 
Thus, COVID-19 can spread differently at the same temperature level. 
Second, the fitted copula models
show that the dependency between the temperature and number of COVID-19 cases was very similar for Riyadh and Jeddah, while it was very low for Mecca. 
Given the first and second points mentioned above, there was another important factor driving the active situation of this new virus, namely, the lockdown situation. 
During the study period, the cities of Jeddah and Mecca were placed under a lockdown on $29$ March 
$2020$ and $30$ March $2020$, 
respectively. Then, Mecca was subjected to a $24$ hour curfew on $02$ April $2020$. Later, on $04$ April $2020$, various areas in Jeddah were placed under a
$24$ hour curfew. After about one day, Riyadh too was placed under a $24$ hour curfew for approximately $18$ days. Then, on $25$ April $2020$, 
Jeddah and Riyadh were placed under partial curfews, while Mecca still remained under a $24$ lockdown. Hence, Mecca experienced a long lockdown 
period, while duration of curfew in Riyadh was the shortest. These factors may explain the similarity in copula results for Riyadh and Jeddah and the low dependency pattern 
in the case of Mecca.
In consideration of these last findings, we can conclude that high temperatures had only a weak to moderate effect on the 
transmission of COVID-19 if there was a partial curfew policy in place. However, this effect vanished under the condition of strong social isolation. Hence, even in hot places, 
COVID-19 can still spread readily when no social distancing is implemented. 
\section*{Conclusion}
This study examined the effect of high temperatures on the spread of COVID-19 in hot climates under different curfew situations using copula models. We applied the models
to the cities of Riyadh, Jeddah, and Mecca in Saudi Arabia. For Riyadh and Jeddah, which had almost the same average temperature level, the association between 
temperature and confirmed cases of COVID-19 reflected a moderate positive Frank copula. However, the number of COVID-19 cases in Riyadh was 
higher than the number in Jeddah. Hence, the transmission of this virus in these two cities may have been affected by the curfew level and not by the
high temperature. In the case of Mecca, which had a temperature level (slightly) less than that of Riyadh and Jeddah, there was a very weak 
dependency between temperature and the number of COVID-19 cases. However, the number of confirmed cases of COVID-19 in Mecca 
was very close to the number in Jeddah. In addition, Mecca was under a strong $24$ hour lockdown for more than half of the observed data set.
Therefore, there is clear evidence that high temperatures are not able to stop the spread of this virus if there is no social isolation. Clearly, 
lockdowns represent the most effective strategy to prevent the spread of this virus. To the best of our knowledge, this study describes the first copula model
fitted to COVID-19 data. The results of this study, derived using copula models, are unlike those derived using existing traditional methods and indicate that the association between COVID-19 and temperature
is weak and no substantial decreases in the number of COVID-19 cases can be expected in response to high temperatures. 

\nolinenumbers

\bibliographystyle{plos2015.bst}

%
%
%
%
%
%
%
%

\end{document}